\documentclass[12pt,preprint]{aastex}
\newcommand{\Msun}{\ifmmode\mbox{M}_{\odot}\else$\mbox{M}_{\odot}$\fi}
\newcommand{\Rsun}{\ifmmode\mbox{R}_{\odot}\else$\mbox{R}_{\odot}$\fi}
\newcommand{\Mearth}{\ifmmode\mbox{M}_{\oplus}\else$\mbox{M}_{\oplus}$\fi}
\newcommand{\Rearth}{\ifmmode\mbox{R}_{\oplus}\else$\mbox{R}_{\oplus}$\fi}

\newcommand{\xmm}{\textit{XMM}\,}

\newcommand{\ntst}{PSR~B1916$+$14\,}

\shorttitle{X-ray Detection of High-B pulsar PSR J1916+14}
\shortauthors{Zhu, Gonzalez, Kaspi \& Lyne}

\begin{document}
\title{XMM-Newton X-ray Detection of the High Magnetic Field Radio Pulsar PSR
B1916+14}
\author{Weiwei Zhu\altaffilmark{1},
Victoria M. Kaspi\altaffilmark{1,2},
Marjorie E. Gonzalez\altaffilmark{3},
Andrew G. Lyne\altaffilmark{4}
\altaffiltext{1}{\footnotesize Department of Physics,
McGill University, Montreal, QC, H3A 2T8, Canada;
zhuww@physics.mcgill.ca, vkaspi@physics.mcgill.ca }
\altaffiltext{2}{Canada Research Chair; Lorne Trottier Chair;
  R. Howard Webster Fellow of CIFAR}
\altaffiltext{3}{Department of Physics and Astronomy, University of British
Columbia, 6224 Agricultural Road Vancouver, BC, V6T 1Z1 Canada; }
\altaffiltext{3}{University of Manchester,
Jodrell Bank Observatory, Macclesfield, Cheshire, SK11 9DL, UK. }
}
\begin{abstract}
Using observations made with the {\it XMM-Newton} Observatory, 
we report the first X-ray detection of the high magnetic field
radio pulsar PSR B1916+14.
We show that the X-ray spectrum of the pulsar can be well fitted with an absorbed
blackbody with temperature in the range of 0.08-0.23 keV, or a
neutron star hydrogen atmosphere model with best-fit effective 
temperature of $\sim$0.10 keV, higher than 
expected from fast cooling models.  The origin of the likely thermal emission is
not well constrained by our short observation and is consistent with initial cooling or
return-current heating.
We found no pulsations in these data and set a 
1$\sigma$ upper limit on the pulsed fraction in the 0.1--2 keV band of
$\sim$0.7.
Implications of these results for our understanding of the different
observational properties of isolated neutron stars are discussed.
\end{abstract}


\section{Introduction}
\label{sec:intr}

In the past few decades, the boom in X-ray/gamma-ray astronomy has led to the 
discoveries of at least two new classes of isolated neutron star:  magnetars and dim
thermal isolated neutron stars (DTINS).
These new classes exhibit distinctive properties different from those of conventional
rotation-powered pulsars,
the only kind of isolated neutron star known before. 
Magnetars, including anomalous X-ray pulsars (AXPs) and soft gamma-ray
repeaters (SGRs), are slowly rotating neutron stars. 
They often emit luminously in the X-ray band and sometimes show dramatic outbursts
and variability.
Their X-ray emission and bursting activity are believed to be powered by
enormous internal magnetic fields ($10^{14}$--$10^{15}$~G; for reviews, see
\citealt{wt06,kas07,me08}).
The DTINS are nearby X-ray pulsars, having quasi-thermal spectra and no 
radio emission. They also appear to be slow rotators and are highly magnetized 
($B\sim10^{13}$~G; \citealt{kv05,zct+05,vk08,kv09}). 
The nature and origin of these DTINS are still mysteries.
A third possible new class is the so-called ``Central Compact Object'' (CCOs,
 \citealt{dmz+08,pl09}), apparently isolated neutron stars located in supernova remnants.
However, it is not yet clear whether these are a truly distinct class as some
CCOs have been recently revealed to be conventional rotation-powered pulsars
(albeit with surprisingly low B fields; e.g. the CCO 1E1207.4$-$5209, 
\citealt{gh07}).

What drives the different behaviors of these neutron stars? 
For magnetars, the answer is almost certainly their enormous magnetic field. 
Interestingly, the DTINS also appear to have higher magnetic fields than
those of ordinary pulsars. 
Hence, it is sensible to suspect that magnetic field could be the primary defining characteristic of the three categories of isolated neutron stars.
Based on the study of a large sample of isolated neutron stars, which
includes most magnetars, some DTINS, and many ordinary radio pulsars,
\citet{plm+07} reported an intriguing correlation between the pulsar's blackbody
temperature $T$, determined from the X-ray spectrum, and the magnetic field $B$,
inferred from the spin-down ($T\propto B^{1/2} $).
They also suggested that this correlation could
be explained if the crusts of neutron stars were heated by magnetic-field decay,
since it would significantly delay the cooling, particularly if the field were stronger than $10^{13}$~G.
A similar correlation was also expected if the core of the neutron star is heated
by magnetic field decay, with heat transfer out to the surface, leading to an
increase of the surface temperature \citep{act04}.
\citet{apm08} expanded the work of \citet{plm+07} through 2D cooling
simulations that included anisotropic thermal conductivity and all relevant
neutrino emission processes for realistic neutron stars, in an attempt to
unite magnetar and DTINS in a simple picture of heating by magnetic field.
Their study shows that the DTINS could be explained if they are old neutron
stars ($\sim 10^6$ yr) born with $10^{14}$--$10^{15}$ G magnetic fields, or if they
are middle-aged neutron stars born with $10^{13}$--$10^{14}$ G magnetic fields.
This theory predicts that pulsars with magnetic field
higher than $10^{13}$~G should be hotter than is predicted by a simple cooling
model with lower magnetic field, regardless of whether they are radio-quiet or
not.
Therefore, observing high magnetic field radio pulsars at X-ray energies and measuring the
temperature of their thermal radiation may help us unify the different classes of isolated
neutron stars.

\ntst\ is a radio pulsar having period $P=$1.181~s, with spin-down-inferred magnetic field
$B\equiv3.2\times10^{19} (P\dot{P})^{1/2}$~G$=1.6\times 10^{13}$~G,
spin-down age $\tau\equiv P/(2\dot{P})=8.8\times10^4$~yr, and
$\dot{E}\equiv 4\pi^2I\dot{P}/P^3=
5\times10^{33}$erg s$^{-1}$ \citep{ht74,atnf}.
It is a relatively young pulsar. 
Given its age, \ntst\ should still be hot enough to be X-ray detectable, according
to a minimal pulsar cooling model, without magnetic-field-decay heating \citep{pgw06}. 
It is also one of the highest-magnetic-field radio pulsars known and may therefore be hotter because of
magnetic-field decay.
This makes \ntst\ a good test subject for neutron star
cooling models, and hence X-ray observations.


\section{Observations and Results}
\label{sec:res}
\ntst\ was observed by the {\it Newton X-ray Multi-Mirror Mission} (\xmm)
observatory \citep{jla+01} on 2008 March 25. 
Both the European Photon Imaging Camera (EPIC) pn \citep{sbd+01} camera and the
EPIC MOS cameras \citep{taa+01} were operating in full window mode with the thin
filter, and with a pointing offset of 1$'$.107.
We analyzed the data taken in this \xmm observation, and found that 
\ntst\ was clearly detected in both the pn and MOS data.

\subsection{Imaging and Source Detection}
\label{sec:imag}
The \xmm data were analyzed with the \xmm Science Analysis System ({\rm SAS}) version
8.0.0\footnote{See http://xmm.esac.esa.int/sas/8.0.0/} and the latest
calibrations (updated 2008 Oct 3).
To exclude strong background flares that sometimes contaminate \xmm data, we extracted
light curves of photons above 10 keV from the entire field-of-view
of the pn and MOS images, and excluded the time intervals in which background flares
occurred for all subsequent analyses.
The total exposure time of the observation is $\sim$ 25 ks. 
However, after excluding the bad time intervals within which the background
flux was very high ($>$10 counts per second) and showing significant burst-like
features, only 12 ks of pn, 11 ks of MOS1 and 13 ks of MOS2
data were used in our analysis.
The data were also corrected to the barycenter using the
{\rm SAS} {\tt barycen} tool after background flares were excluded, using the
nominal pulsar position (J2000) RA 19:18:23.638(7)\, DEC +14:45:06.00(15) (\citealt{hlk+04}).

\begin{figure}
\includegraphics[scale=0.6]{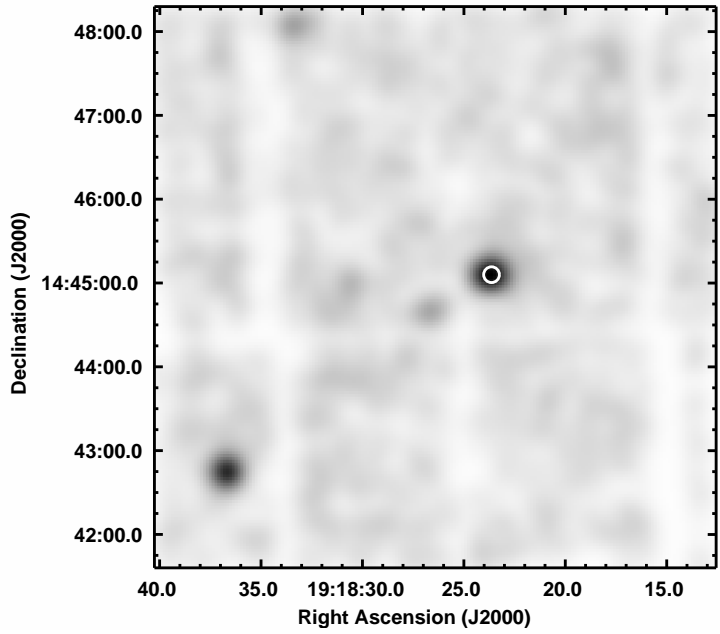} \\ 
\caption {\label{fig:image} \xmm image in the 0.2--2 keV energy band, smoothed by a Gaussian profile of
8.2$''$ radius (a profile that is slightly oversampling the telescope's PSF). The radio position of \ntst\ is labeled by the white circle.
Note that the radius of the circle is much larger than the uncertainty on the
radio position.} 
\end{figure}

In order to find the X-ray counterpart of \ntst, we used the {\rm SAS} tool
{\tt edetect\_chain} to perform a blind search for point sources. 
{\tt edetect\_chain} is designed to find point sources using a sliding
cell technique and to calculate the significance of any detection using a maximum likelihood method. 
It generates an output source list file containing information like total counts, position
and significance of detected sources. 
In the pn image, a point source was detected coincident with the position of
\ntst\ (Fig. \ref{fig:image}) by {\tt edetect\_chain}. 
The source has $133\pm15$\footnote{Unless otherwise specified, the uncertainties quoted in this paper represent a
1$\sigma$ range.} counts in 0.1--10 keV band and a likelihood ratio of $L_2=-\ln(P)\simeq 143$ (where $P$ is the probability for a random Poissonian fluctuation to have caused the observed source counts).
This source was also detected in the MOS 1 image with $22\pm7$ counts and
$L_2\simeq 11$, and in the MOS 2 images with $46\pm9$ counts and $L_2\simeq
48$, both in the 0.1--10 keV range. 
Thus, this point source was clearly detected in the pn and the MOS data.

Figure \ref{fig:image} is the pn image, smoothed with a Gaussian profile of
radius $\sigma=8''.2$. 
The small circle in the center of the image marks the radio position of
\ntst. The position uncertainty is smaller
than the size of the circle \citep{hlk+04}. 
The best-fit position of the detected source given by {\tt edetect\_chain} is (J2000) RA 19:18:23.74(5)\, 
DEC +14:45:06.2(8), consistent with the radio position of \ntst.
Therefore, it is very likely that the source we detected is the X-ray counterpart of
\ntst.

The radially averaged profile of \xmm's point spread
function can be approximated by an analytic function -- the King function
$\rho(r)=A[1+(\frac{r}{r_0})^2]^{-\alpha}$, where $\rho(r)$ is the number
density of counts at radius $r$, $A$ is a normalization parameter,
$r$ is the radial distance between the events and the center of the source,
$r_0$ and $\alpha$ are parameters reflecting the size and shape of the point
spread function (PSF)
and are functions of energy and off-axis angle.\footnote{See
http://xmm.esac.esa.int/docs/documents/CAL-TN-0018-2-6.pdf, page 6}  
In order to search for evidence of extended emission from \ntst, we
extracted 0.2-12 keV photon events from a circular region of 35$''$ radius
around the best-fit position, and calculated the radial distance between every photon event and
the pulsar, to get the radially averaged profile. Using the Kolmogorov-Smirnov (K-S) test,
we then compared  the observed radially averaged profile to a model composed
of a King function. Given the small off-axis angle, and the energy distribution of
the source, we chose $\alpha=1.6$ and $r_0=5.25$ pixels$=21''.525$ and a
uniform background (0.05 photons/acrsec$^2$, inferred from the 397 photons
found in a circular background region of 50$''$ radius and $\sim$3$'$
away from \ntst). The K-S test shows that the radially averaged
profile can be well matched by the specified King function. Therefore, there
is no evidence for extended emission near \ntst\, from this observation.

\subsection{Spectroscopy}
\label{sec:spec}
We extracted the X-ray spectrum of \ntst\ from the pn data using a circular
region of 32$''$.5 radius encircling the source. 
The source region should contain more than 80\% of the counts from a point
source.
The background spectrum was extracted from a circular region of 50$''$ radius and
$\sim$3$'$ away from the pulsar where no source was detected.
Both single- and double- events were selected, but events that hit or were close
to a bad pixel or CCD gap were excluded using the filter expression $\rm FLAG=0\&\&PATTERN <=4$.
A response file and an auxiliary response file were generated using the
{\rm SAS} command {\tt rmfgen} and {\tt arfgen}.
The spectrum was grouped to have a minimum of 15 photons per bin using the {\tt
ftool} {\tt grppha}, and was then fed to {\rm XSPEC
12.3.0}\footnote{http://heasarc.gsfc.nasa.gov/docs/xanadu/xspec/} for spectral fitting.

We also extracted spectra from the data of the two MOS
detectors using source circular regions of 36$''$ radius and background
regions of $\sim60''$ radius. 
Single- to quadruple- photon events were selected except those that landed on a
bad pixel or CCD gap, using the filter
expression of $\rm XMMEA\_EM \&\& PATTERN <=12 $. 
We then combined the two MOS spectra into a single MOS spectrum and averaged their
background, response and auxiliary files using the {\tt ftool} {\tt addspec}. 
The resulting MOS spectrum was also grouped to have a minimum of 15 photons per bin and
was fitted jointly with the pn spectrum.

\begin{figure}
\includegraphics[scale=0.6,angle=270]{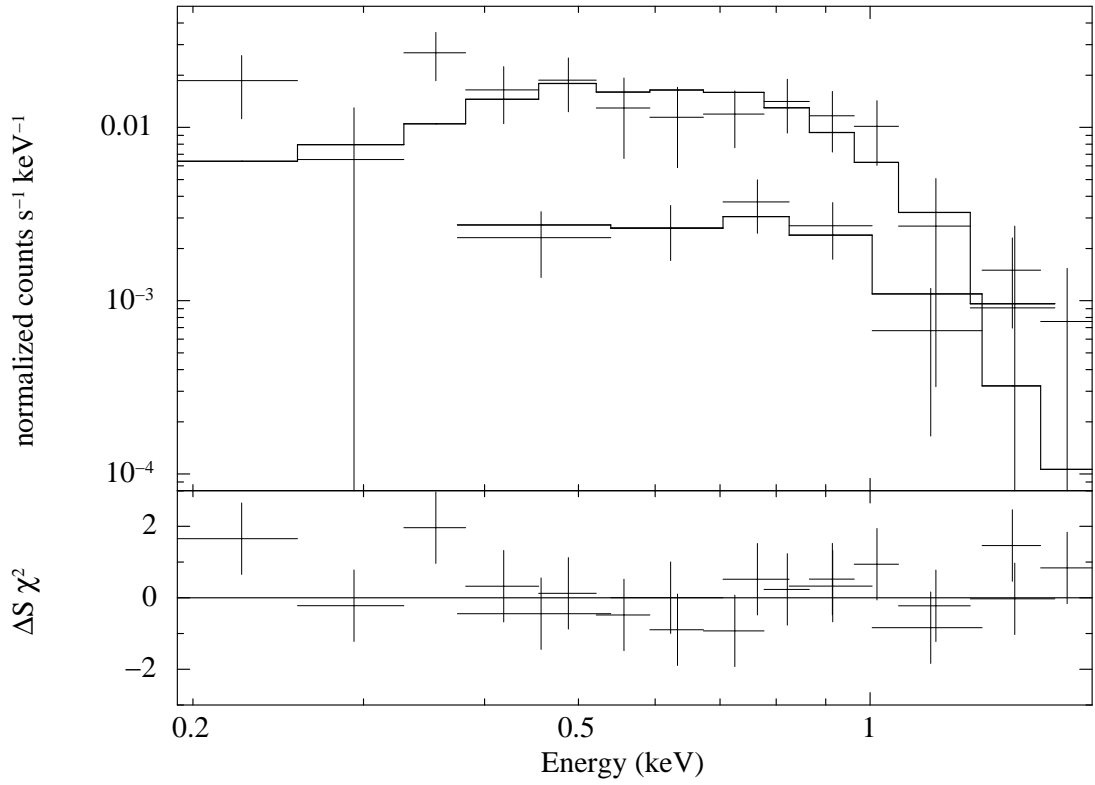} \\ 
\caption {\label{fig:spec} \xmm spectra (upper is pn, 
lower is combined MOS) of \ntst, with the best blackbody fit
(see Table \ref{tabSpecFit}). The spectra are binned to contain a minimum of
15 counts per bin. } 
\end{figure}

The X-ray spectra of \ntst\ can be well fit with an absorbed blackbody
model. 
However, due to the small number of counts, the column density $N_H$ was
poorly constrained.
The best-fit $N_H$ is $\sim1\times10^{19}$~cm$^{-2}$, too small given the estimated distance and location of the pulsar.
Therefore, we estimated the $N_H$ for this pulsar based on the total $N_H$
(1.58$\times 10^{22}$~cm$^{-2}$) of the
Galaxy along the
line-of-sight\footnote{http://cxc.harvard.edu/toolkit/colden.jsp}
and the distance to the pulsar
($2.1\pm0.3$~kpc, estimated from the 27.2~pc~cm$^{-3}$ dispersion measure of the pulsar; \citealp{cl02}), and find a moderate value of $\sim0.14\times10^{22}$~cm$^{-2}$.
Fixing $N_H$ to this value, the best-fit blackbody temperature for the
0.1--2 keV spectra (Fig. \ref{fig:spec}) is $0.13\pm 0.01$ keV
(Table \ref{tabSpecFit}).
The model-predicted absorbed flux in the 0.1--2 keV range is
1.4$\pm0.3\times$10$^{-14}$~ergs~s$^{-1}$~cm$^{-2}$.
Assuming a distance of 2.1 kpc,
we find the bolometric X-ray luminosity of \ntst~ to be $\sim
3\times10^{31}$~ergs s$^{-1}$.

By fixing $N_H$ while fitting the pn and MOS spectra, we underestimate the
uncertainties of the best-fit parameters.
In order to get a sense of the real uncertainty of $kT$, we tried fitting with a range of
$N_H$.
$N_H$ likely lies between $0.07\times10^{22}$~cm$^{-2}$ and $0.3\times10^{22}$~cm$^{-2}$.
It is probably not smaller than $0.07\times10^{22}$~cm$^{-2}$ because the
distance estimated from the dispersion measure is unlikely to be incorrect by more than 50\%.
Also, an absorbed blackbody model with $N_H$ higher than
$0.3\times10^{22}$~cm$^{-2}$ cannot fit the spectra well for any $kT$.
With $N_H$ restricted to lie between these two values, the acceptable (null
hypothesis possibility of the fit $>$ 0.01) range of
$kT$ is 0.08 to 0.23 keV (with blackbody radius range from $\sim$6 km to
$\sim$0.2 km).
This temperature range reflects reasonable uncertainties on $N_H$, and is
quoted in the abstract and Figures \ref{fig:agekt} and \ref{fig:bkt} (see below).

The pn and MOS spectra could also be fit with a power-law model, with
 a best-fit $N_H$ of $0.12^{+0.05}_{-0.07}\times10^{22}$ $\rm cm^{-2}$ and a photon
index of $\Gamma\sim$3.5$^{+1.6}_{-0.7}$ (Table \ref{tabSpecFit}).
The lack of source photons with energy above 2 keV results in a soft
best-fit power-law model. 
This is rarely seen from other non-thermally emitting pulsars.
Therefore, it is very unlikely that the X-ray emission of \ntst\ is
non-thermal.

A neutron-star hydrogen atmosphere (NSA) model (with magnetic field strength
set to $10^{13}$ G; \citealp{zps96,pszm94}) could also fit the
pn and MOS spectra. However, the parameters are even less constrained in comparison with the
blackbody model. We had to freeze the mass of the neutron star to 1.4 $M_{\odot}$ 
and the distance to 2.1 kpc to get a better-constrained fit ($\chi^2(\nu)=14.3(17)$). The best-fit
$N_H$ is $0.23^{+0.09}_{-0.04}\times10^{22}$cm$^{-2}$, $kT$ is $0.10\pm0.04$ keV, and the
resulting best-fit neutron-star radius is $\sim$6 km. 
Unfortunately, the radius is not well constrained in this model and has a 1$\sigma$ upper limit
of 20 km.
The model predicted 0.1--2 keV X-ray luminosity is
$(7\pm4)\times10^{31}$ergs~s$^{-1}$.

\subsection{Timing analysis}
\label{sec:pf}
To search for X-ray pulsations from \ntst, we folded all the pn source events
from a total 25 ks exposure without filtering for the background flares 
using 16 phase bins and a contemporaneous ephemeris which was derived
from radio timing data obtained using the 76-m telescope at the Jodrell Bank
Observatory \citep{hlk+04}.  
The MOS full-window mode data is useless
for timing analysis because of its 2.7 s time resolution.
A total of 374 pn photon events were extracted without filtering for background flares, all in the range of 0.1-2 keV, from a source
region of 15$''$ radius chosen to reduce the number of background photons and improve
the signal-to-noise ratio.
From the same energy band and the same CCD, 1945 events were found in a circular background
region of 50$''$ radius where no source was detected by the {\rm SAS} tool {\tt edetect\_chain}. 
If the background is uniformly distributed, there should be 175$\pm$4
background photons in the source region.
The folded and binned light curve was fit to a constant line. 
The best-fit $\chi^2$ was 9.6 for 15 degrees of freedom. 
Therefore no significant pulsations were detected.

In order to determine an upper limit on the pulsed fraction, we
simulated event lists with the same total number of counts as in the
observed event list. 
The simulated event lists were generated assuming the signal has a sinusoidal
profile starting at a random phase and a specified area pulsed fraction, 
where the area pulsed fraction is defined as the ratio of the pulsed part of profile to the entire profile. 
For a sinusoidal profile $F=A\sin\theta+B$, where $F$ is the
count rate, $A$ is the amplitude, $B$ is the DC level, and $\theta$ is the
phase, the area pulsed fraction is $A/(2B+A)$.  
By specifying different area pulsed fractions, we
found that, if we set the area pulsed fraction of the simulated event lists to
35\%, then approximately 68\% of them would be detected with $>3\sigma$ significance.
Because there are $\sim$175 background photons in the 374 photons from the source
region, the 1$\sigma$ area pulsed fraction upper limit of the pulsar is $\sim0.7$.
This is not an especially interesting constraint because the number of source photons was so small that even a highly pulsed signal could have gone undetected.

\section{Discussion}
\label{sec:disc}

The X-ray spectrum of \ntst\, is soft, and therefore is most likely thermal.
The blackbody temperature lies in the range of 0.08--0.23 keV, and the
best-fit effective temperature of  NSA model is $0.10\pm0.04$ keV.
Given \ntst's age, its temperature is
consistent with what one would expect ($kT\sim0.07$--$0.11$ keV) from minimal cooling models in which
magnetic field is not considered \citep{pgw06}.  
Its estimated bolometric luminosity ($\sim 3\times10^{31}$erg~s$^{-1}$; Table
\ref{tabSpecFit}) is somewhat low when compared with the curves of Page et al. (2006), suggesting that
the pulsar may have a light-element envelope.   However the substantial uncertainties on the luminosity
preclude a firm conclusion.  On the other hand, fast cooling models predict much lower temperatures
($kT<30$ eV; e.g. Yakovlev \& Pethick 2004) which would be undetectable with current instruments.
Therefore, fast cooling seems unlikely for \ntst\ from our observation.

The small best-fit blackbody radius ($0.8\pm0.1$ km; Table \ref{tabSpecFit}) suggests polar-cap
reheating.
If the emission of the pulsar is due to curvature radiation, return-current
heating is predicted to give rise to an X-ray luminosity (see \citealt{hm01a}
and Eq. 7.2 in \citealt{krh06} for details)
\begin{equation}
\label{xlum}
L^{(CR)}_+\simeq10^{31} {\rm erg~s^{-1}} \left\{\begin{array}{ll}
0.4P^{-6/7}(\tau/10^6)^{-1/7} & {\rm if}\, P\le 0.1(B/10^{12})^{4/9}\\
1.0P^{-1/2} & {\rm if}\, P\ge 0.1(B/10^{12})^{4/9}, \\
\end{array}\right.
\end{equation}
where $\tau$ is in yr, $P$ is in s, and $B$ is in G.
For \ntst, $L^{(CR)}_+$ is $\sim9\times10^{30}$erg~s$^{-1}$, only a factor of
$\sim$3 smaller than the estimated blackbody bolometric luminosity (see Table
\ref{tabSpecFit}).
If the emission is due to inverse Compton scattering, the return
current heating will be much less effective \citep{hm02}. 
Given that the best-fit blackbody parameters are not well constrained, we cannot
rule out return current heating as the origin of \ntst's X-ray luminosity.
We note that the NSA model (Section \ref{sec:spec}) yields a larger
radius, although it is also not well constrained.

\begin{figure}
\includegraphics[scale=0.6]{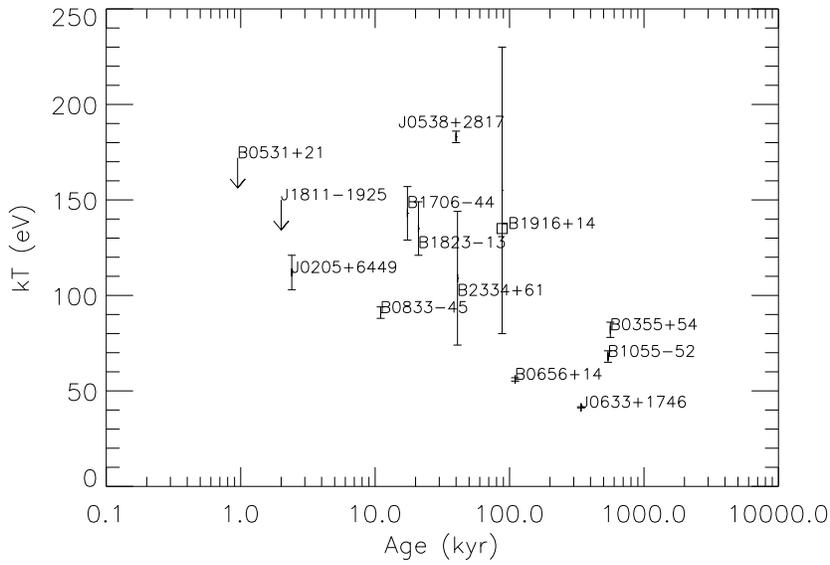} \\ 
\caption {\label{fig:agekt} Observed temperature ($kT$) versus age for
X-ray-detected radio pulsars. Note that for \ntst\ we used the $kT$ measured by
allowing $N_H$ to vary in a reasonable range; the same is not necessarily true
for the other measurements; see original references for details.}
\end{figure}

\begin{figure}
\includegraphics[scale=0.6]{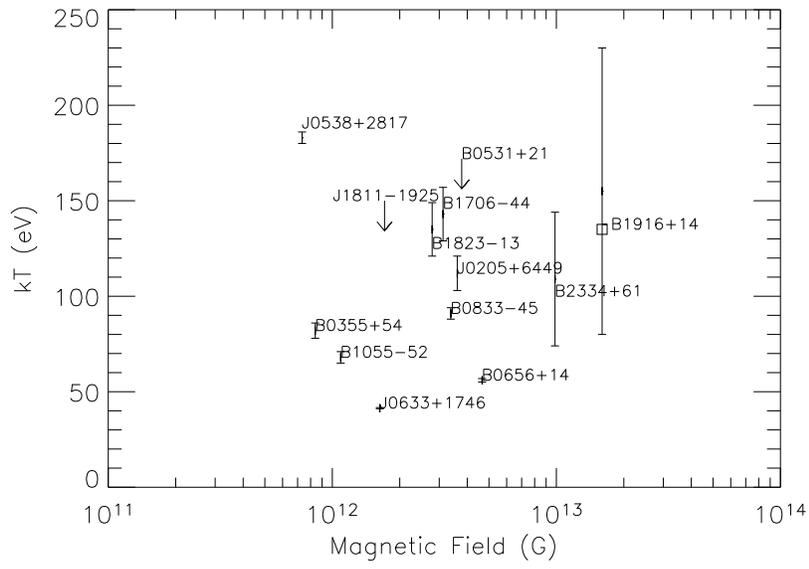} \\ 
\caption { \label{fig:bkt} Observed temperature ($kT$) versus magnetic field
strength for X-ray-detected radio pulsars. See caption for Figure
\ref{fig:agekt} for caveats.}  
\end{figure}

To compare the properties of \ntst\ to those of other X-ray-detected radio
pulsars, we have collected the temperature, magnetic field strength, and spin-down
energy of a dozen such pulsars from the literature, and listed them in Table
\ref{tabPSRs}.
For pulsars in this Table, we also made plots of 
their temperature versus age and magnetic field 
(Figs. \ref{fig:agekt} and \ref{fig:bkt}, respectively).
Given the large uncertainty on the temperature measurement from our short-exposure
observation of \ntst, we cannot conclude here whether thermal emission is
consistent with minimal cooling or if the neutron star is hotter than
lower-magnetic-field pulsars of the same age, as expected in some models
\citep{plm+07,apm08}.
A longer observation in the future may be able to
distinguish among thermal models.

However, we note with interest that the previously published temperature of PSR J0538+2817\, is
surprisingly high, in spite of its relatively large age (40 kyr) and relatively low magnetic
field ($7.3\times10^{11}$ G; Table \ref{tabPSRs}).  In contrast to \ntst,
PSR~J0538+2817's emission is unlikely to be from polar-cap reheating because of
the very high required efficiency for conversion of $\dot{E}$ to X-ray
luminosity, $\sim 10^{-2}$, compared with the $\sim 5 \times 10^{-4}$ predicted
by Equation~1 above for this pulsar. 
If correct, the high temperature suggests a wider range of possible temperatures for young
neutron stars than is currently predicted.
This would be a challenge to the \citet{plm+07} model.

Although \ntst\ is detected by \xmm as a point source with no evidence of
extended emission, it is still possible that extended emission was too
faint to be detectable.
Based upon the number of counts in the background region, we found that, in the
0.3--8 keV band, extended emission of surface brightness smaller than
$\sim3\times10^{-6}$~count~s$^{-1}$arcsec$^{-2}$ would not be detected (with
3$\sigma$ significance) in our observation. 
This limits our sensitivity for detecting a very faint pulsar wind nebula (PWN)
like that of the Geminga pulsar, which has a surface brightness of
$\sim1\times10^{-6}$~count~s$^{-1}$arcsec$^{-2}$  \citep{psz06} in the same energy range, despite
the fact that Geminga is closer by a factor of $\sim$8. 
Assuming there is an undetected PWN around \ntst~ having a spectrum like that of the Geminga PWN
(power law with index 1.0), we can estimate the upper limit of its surface
brightness to be $\sim3\times10^{-17}$~erg~cm$^{-2}$s$^{-1}$arcsec$^{-2}$, in the 0.3--8
keV range.
If we further assume that the PWN is uniformly distributed in a circular region
of radius 20$''$, then this surface brightness upper limit corresponds to a
PWN luminosity upper limit of
$\sim$3$\times10^{31}$~erg~s$^{-1}$.

\citet{gkp04} showed that the presence of a very high magnetic field could cause
inhomogeneous thermal conductivity in the neutron star crust and lead to the
formation of hot spots on the neutron star surface. It could also cause strong
radiative beaming of the thermal emission from the neutron star. These
effects could give rise to highly pulsed X-rays, as in the
$74\pm14\%$ pulsed fraction observed from the high-magnetic-field radio
pulsar PSR J1119$-$6127 \citep{gkc+05}. 
However, limited by the exposure time of the observation, we did not detect any
X-ray pulsations from \ntst.
Future longer observations will be useful for better constraining its pulsed
fraction.

\acknowledgements
We thank S. Bogdanov and S. Guillot for useful comments.
We thank the referee Craig Heinke for helpful suggestions.
VMK is supported by an NSERC Discovery Grant, CIFAR,
FQRNT, the Canada Research Chairs Program, and the McGill
Lorne Trottier Chair.


\clearpage

\begin{deluxetable}{lccc}
\tabletypesize{\footnotesize}
\tablewidth{0pt}
\tablecaption{\label{tabSpecFit} Best-fit Spectral Parameters for \ntst 
}%
\tablehead{ \colhead{Parameter} &\colhead{Blackbody model\tablenotemark{a}} &
\colhead{Power-law model} & \colhead{NSA model}}
\startdata
$N_{H}$ (10$^{22}$~cm$^{-2}$) & 0.14 & $0.12^{+0.05}_{-0.07}$ & $0.23^{+0.09}_{-0.04}$\\
$kT$ (keV)  & 0.13$\pm 0.01$ & --- & $0.10\pm0.04$\\
$R_{bb}$ (km) & $0.8\pm0.1 $& --- & ---\\
$R_{NS}$ (km) & --- & --- & $\sim$6\\
$\Gamma$ & ---  & 3.5$_{-0.7}^{+1.6}$ & ---\\
$\chi^2$(dof)  & 14.1(18) & 13.9(17) & 14.3(17)\\
$f_{\rm abs}$\tablenotemark{b} (ergs~s$^{-1}$~cm$^{-2}$) &
1.4$\pm0.3\times$10$^{-14}$ & 1.7$^{+0.4}_{-0.6}$$\times$10$^{-14}$&
(1.7$\pm0.3)\times10^{-14}$\tablenotemark{d}\\
$f_{\rm unabs}$\tablenotemark{c} (ergs~s$^{-1}$~cm$^{-2}$)  &
$(5\pm1)\times$10$^{-14}$ & $\sim2\times$10$^{-13}$ &
 (1.4$\pm0.8)\times10^{-13}$\tablenotemark{d}\\
$L_{\rm X}$  (ergs~s$^{-1}$)  &
$\sim3\times$10$^{31}$\tablenotemark{e} &
$\sim1\times$10$^{32}$\tablenotemark{f} &
$(7\pm4)\times10^{31} $\tablenotemark{d}\\
\enddata
\tablenotetext{a}{Best-fit parameters of absorbed blackbody fit to the \xmm
spectra. $N_H$ was frozen when fitting, so the uncertainties of the
parameters, especially that of the $kT$, do not reflect the uncertainties on
$N_H$; see text for details. Emission radius $R_{bb}$ was inferred assuming a distance of $2.1$~kpc (estimated from the dispersion measure; \citealt{hlk+04}).}
\tablenotetext{b}{Absorbed X-ray flux, $f_{\rm abs}$ in the 0.1--2 keV range.}
\tablenotetext{c}{Unabsorbed X-ray flux, 
$f_{\rm unabs}$, in the 0.1--2 keV range; the uncertainty was propagated
from the uncertainties on the parameters and absorbed flux.}
\tablenotetext{d}{When fitting with NSA model, the flux and luminosity are estimated using the {\tt cflux} model in
{\tt xspec}.}
\tablenotetext{e}{Bolometric X-ray luminosity derived
assuming a distance of 2.1~kpc.}
\tablenotetext{f}{Inferred X-ray luminosity in the 0.1--10 keV range assuming
a distance of 2.1~kpc.}

\end{deluxetable}

\clearpage

\begin{deluxetable}{lccllccc}
\rotate{}
\tabletypesize{\footnotesize}
\tablewidth{0pt}
\tablecaption{\label{tabPSRs} Parameters of the X-ray-detected Radio Pulsars 
}%
\tablehead{ \colhead{Name} & \colhead{Age\tablenotemark{a} (kyr)} & \colhead{$kT$\tablenotemark{b} (eV)}  & \colhead{$B$\tablenotemark{c} (G)} & \colhead{$\dot{E}$\tablenotemark{c} (erg\,s$^{-1}$)} &\colhead{Observatory} & \colhead{Spectral model\tablenotemark{d}} & \colhead{Reference} }
\startdata

B1916+14&	88&	135(45)&  1.6$\times10^{13}$&   $5.1\times10^{33}$& {\it XMM}& BB& this work\\
B1055$-$52&	540&	68(3)&	  1.1$\times10^{12}$&  $3.0\times10^{34}$&	{\it XMM}& BB+BB+PL& \citealt{dcm+05}\\
J0633+1746 (Geminga)&	340&	41.4(0.1)&1.6$\times10^{12}$&  $3.2\times10^{34}$&	{\it XMM}& BB+PL& 	\citealt{jh05}\\
B0656+14&	110&	56.0(0.9)&4.7$\times10^{12}$&  $3.8\times10^{34}$&	{\it XMM}& BB+BB+PL& \citealt{dcm+05}\\
B0355+54&	562&	82(4)&	  8.4$\times10^{11}$&  $4.5\times10^{34}$& {\it ROSAT/Einstein} & BB& \citealt{sla94}\\
J0538+2817&	40\tablenotemark{e}&	183(3)&	  7.3$\times10^{11}$&  $4.9\times10^{34}$&	{\it XMM}& BB&	 \citealt{mkz+03}\\
B2334+61&	41&	109(35)&  9.9$\times10^{12}$&  $6.2\times10^{34}$&	{\it XMM}& BB+PL& \citealt{mzc+06}\\
B1823$-$13&	21&	135(14)&  2.3$\times10^{12}$&  $2.8\times10^{36}$&	{\it Chandra}& BB&	\citealt{pkb08}\\
B1706$-$44&	17.4&	143(14)&  3.1$\times10^{12}$&  $3.4\times10^{36}$&	{\it Chandra}& BB+PL& \citealt{ghd02}\\
J1811$-$1925&	2\tablenotemark{f}&	$<$150&	  1.7$\times10^{12}$&
$6.4\times10^{36}$&	{\it Chandra}& BB+PL&	\citealt{krh06}\\
B0833$-$45 (Vela)&	11&	91(3)&	  3.4$\times10^{12}$&  $6.9\times10^{36}$&	{\it XMM}& BB+BB+PL& \citealt{mdc07}\\
J0205+6449&	2.4\tablenotemark{f}&	112(9)&	  3.6$\times10^{12}$&  $2.7\times10^{37}$&	{\it Chandra}& BB+PL& \citealt{shvm04}\\
B0531+21 (Crab)&	0.955\tablenotemark{f}&	$<$172&	  3.8$\times10^{12}$&  $4.6\times10^{38}$&	{\it Chandra}& BB+PL&	\citealt{wop+04}\\

\enddata
\tablenotetext{a}{The spin-down age unless otherwise noted.}
\tablenotetext{b}{The blackbody temperature or the temperature of the softer
blackbody component as measured by fitting the data with different spectral
models as listed in this Table.}
\tablenotetext{c}{Numbers were found in the ATNF database \citep{atnf}.}
\tablenotetext{d}{BB: blackbody model; BB+PL: blackbody plus power-law model;
BB+BB+PL: two blackbody plus power-law model.}
\tablenotetext{e}{The age of PSR J0538+2817~ estimated based on the proper motion of the
pulsar from its associated supernova remnant (SNR) \citep{nrb+07}.}
\tablenotetext{f}{The age of SNR with which the pulsar is associated,
estimated based on its expansion rate: PSR J1811$-$1925 \citep{krh06}; PSR
J0205+6449  \citep{che05}. The age of PSR B0531+21 (Crab pulsar) is based on
historical record.}
\end{deluxetable}

\clearpage

\clearpage

\clearpage

\clearpage

\end{document}